\documentclass{epsconf}
\usepackage{graphicx}
\usepackage{wrapfig}
\usepackage{amsmath}

\title{Development of LPA based hard X-ray sources at ELI Beamlines}
\author{U. Chaulagain$^1$, M. Lamač$^{1,2}$, M. Raclavsky$^{1,3}$, K. H. Rao$^1$,  J. Nejdl$^{1,3}$, S. A Weber$^1$, and S. V. Bulanov$^1$}
\institute{$^1$ ELI Beamlines, Institute of Physics ASCR, Prague, Czech Republic\\
$^2$ FMP, Charles University, Prague, Czech Republic\\
$^3$ FNSPE, Czech Technical University in Prague, Prague, Czech Republic}

\begin{document}
\maketitle
\section{Introduction}
 ELI-Beamlines is a part of the ELI (Extreme Light Infrastructure) project and will soon become one of the most powerful laser facilities in the world \cite{Rus2011}. One of the main objectives of ELI Beamlines center, is to provide beams of laser-driven femtoseconds pulses of hard X-ray radiation to the users from various fields of research \cite{Nejdl2020}. Based on radiation from electrons accelerated in plasma, the Betatron and Inverse Compton X-ray sources have the common properties to be compact and to deliver collimated, incoherent, and femtosecond radiation. The Betatron X-ray source is laser-driven broadband hard X-ray source, generated from the transverse oscillations of electrons during their acceleration and propagated along the direction of electron velocity. It is characterized with a source size of the order of a micrometer, a pulse duration of the order of few femtoseconds. These unique features of the source will be open to various applications in atomic physics to biomolecular sciences as well as in the industry. Here we report progress on development of these noval X-ray sources at ELI Beamlines facility. We also report on an advanced interferometry method for characterization of gas jets that are routinely used in laser-plasma accleratror (LPA) as a target.

\section{LPA based X-ray sources at ELI beamlines}
ELI Gammatron beamline \cite{Chaulagain2022}  is multidisciplinary, user-oriented hard X-ray sources of radiation based on the Laser Wakefield acceleration \cite{Tajima1979}. It provides X-ray pulses with a photon flux beyond $10^{11}$ photons per pulse with  optimized laser and plasma conditions \cite{Kozlova2020,Fourmaux2020,Lamac2021}. The X-ray pulses can be either broadband or quasi-monoenergetic with photon energies ranging from a few keV up to a few MeV based on Betatron radiation or Inverse Compton radiation. These X-ray sources have pulse duration of few femtoseconds as they inherit the temporal properties of the accelerated electron bunch. 

\begin{figure}[h]
	\centering
	\includegraphics[width=16cm]{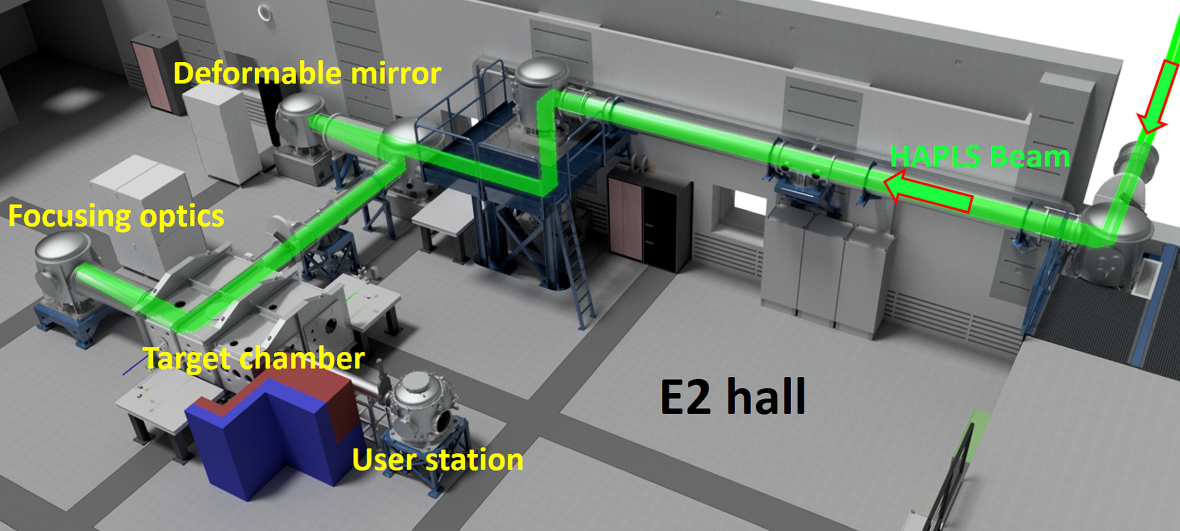}
	\caption{ELI Gammatron beamline located in the E2 experimental hall of ELI-Beamlines.}
	\label{epsfig1}
\end{figure}

Fig. \ref{epsfig1} shows the Gammatron beamline located at the experimnetal hall E2. The driving laser (\textit{Green ~beam}) is the state-of-the-art diode-pumped Ti:sapphire HAPLS laser (30J, 10 Hz). The beam is guided from the laser compressor to the interaction chamber using six turning box chambers equipped with  high laser damage thresold ion beam sputtered dilectric coated mirrors \cite{Willemsen2022}. The laser beam focus to supersonic gas jets drives a LPA in which electrons accelerate to GeV level and produce ultrashort pulse of hard X-ray radiation. The gas jet target allows to operate the source at high-repetition rate, typical gases used are pure He, mixture of He with Nitrogen, or even simply dry air\cite{Bohacek2018}. The all-optical ultrafast X-ray pulse from the Gammatron beamline, with jitter-free synchronization with the pump IR beam, due to their way of generation from the high-power optical laser pulses, provides an extensive platform to perform time-resolved ultrafast studies on atomic time scales. A dedicated user station has been designed to enable various time-resolved experiments. We have designed a novel multi-lane KB X-ray focusing optics covering a wide spectral range up to 25 keV which aims to exploit the broadband nature of the betatron X-ray source for different applications ranging from phase contrast imaging \cite{Chaulagain2017} to time-resolved X-ray spectroscopy and diffraction of bulk samples.

In addition, a high peak brightness hard X-ray Betatron source is being commissioned within the plasma physics platform (P3) in the E3 experimental hall \cite{Weber2017,Chaulagain2018}. The P3 aims to carry wide range of advanced plasma physics experiments including HED physics \cite{Vidal2017}, laboratory astrophysics \cite{Chaulagain2015,Singh2017} and high field interactions. These experiments are very important, in particular, to benchmark the models and simulation codes that are often used to model the astrophysical process \cite{Cotelo2015,Vidal2021}. The P3 platform is designed to perform advanced plasma physics reserach using multiple lasers beams; with both the high energy long pulse and high power short pulse lasers \cite{Jourdain2021}. Presence of a hard X-ray backlighter adds the capability of P3  in diagnostics of highly ionized dense plasmas genreated during high-intensity laser matter interaction. 

\section{Advanced interferometry  target characterization station for LPA targets}
\begin{figure}[h]
	\centering
	\includegraphics[width=16cm]{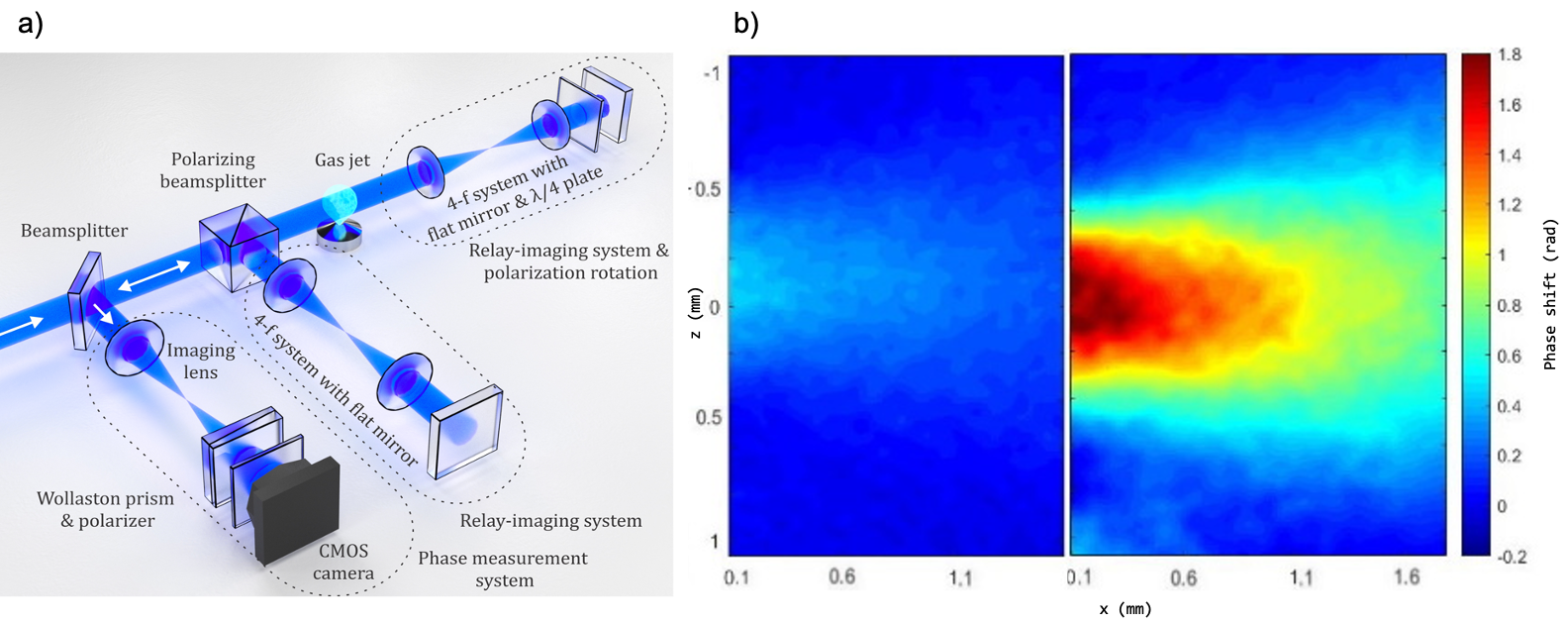}
	\caption{a) Schematics of four pass set up for interferrometric gas jet characterization.  b) Comparison of spatial phase distributions obtained at one pass (\textit{left}), and four passes (\textit{right}) of the probe beam through the gas jet generated by a 1-mm cylindrical nozzle (P= 3 bar, Ar gas). }
	\label{epsfig2}
\end{figure}

We have developed a novel optical probing technique with significantly increased sensitivity and high spatial resolution suitable for characterization of low-density LPA gas  targets. These features are a result of unique optical configuration employing multiple passes of the probe beam through the object with two relay-imaging arms and polarization switching using passive optical elements. With  multiple passes of the probe beam through the sample to enable an increase of the accumulated phase and improve the sensitivity. The scheme of four pass interferrometry is shown in \ref{epsfig2}a. In a two-pass configuration a two-fold increase in the sensitivity compared to the standard single-pass configuration  \cite{Nejdl2019, Chaulagain2020} and with four-fold in four pass was found. Fig. \ref{epsfig2}b shows the comaprision of the measurment at single pass to four pass configuartion, we observed a four-fold increase in the phase sensitivity. We used this advanced interferometric technique to perform a tomographic characterization of slit nozzle with complex spatial distribution of the gas density \cite{Karatodorov2021}.

\section{Acknowledgments} Supported by the project Advanced research using high intensity laser produced photons and particles (ADONIS) (CZ.02.1.01/0.0/0.0/16\_019/0000789) from European Regional Development Fund and Project LM2018141 of the Ministry of Education, Youth and Sports within targeted support of Large infrastructures.

\end{document}